# Semi-Parallel Deep Neural Networks (SPDNN), Convergence and Generalization

*Shabab Bazrafkan, Peter Corcoran*
*Center for Cognitive, Connected & Computational Imaging, College of Engineering & Informatics, NUI Galway, Galway, Ireland*
*{s.bazrafkan1,peter.corcoran}@nuigalway.ie*

## Introduction:

In recent years the Deep Neural Networks (DNN) has been using widely in a big range of machine learning and data-mining purposes. This pattern recognition approach can handle highly nonlinear problems due to their nonlinear activation function of each neuron which can reorient the input data in their most representative form based on the output values. These networks are made of several signal processing parts placed consecutively. In image understanding tasks these signal processing parts usually are convolutional layers, fully connected layers, accompanied by pooling, and regularization tasks.

**Convolutional Layer**: This layer convolves the (in general 3D) image "I" with (in general 4D) kernel "W" and adds an (in general 3D) bias term "b" to it. The output is given by:

$$P = I * W + b$$

Where the $*$ operator is nD convolution in general. In the training process, the kernel and bias parameters are selected in a way to optimize the error function of the network output.

**Pooling Layer**: The pooling layers apply a (usually) non-linear transform (Note that the average pooling is a linear transform, but the more popular max-pooling operation is non-linear) on the input image which reduces the special size of the data representation after the operation. It's common to put a pooling layer between two consecutive convolutional layers. Reducing the spatial size will lead to less computational load and also prevents the over-fitting and also adds a certain amount of translation invariance to the problem.

**Fully Connected Layer**: Fully connected layers are exactly same as the classical Neural Network (NN) layers where all the neurons in a layer are connected to all the neurons in their subsequent layer. The neurons give the summation of their input multiplied by their weights and then passed through their activation functions.

**Regularization**: In general, regularization is proposed to prevent the overfitting inside the network. One can train a more complex network (more parameters) with regularization and prevent over-fitting while the same network would get over-fitted without regularization. Different kind of regularizations has been proposed. The most important ones are weight regularization, drop-out technique, and batch normalization layer. Each of them has their own advantages and drawback which make each one more suitable for specific applications.

In the next section the idea of the Semi-Parallel Deep Neural Networks is presented an example and at the following section described the differences of the proposed method with similar ideas. It has been shown that the SPDNN gives an architecture with the same number of parameters while having a high convergence and improved generalization.

# Semi-Parallel Deep Neural Networks

For one specific problem, one can find a bunch of different architecture of deep neural networks in literature, each of them has their own advantages and drawbacks, from the other point of view there has been a large effort on making the network deeper and deeper to have a more representative reorientation of the input data. Bigger networks always come with two major drawbacks:
:
1. The most important problem is that a bigger network will have more parameters to learn which makes these networks more susceptible to overfitting. There have been big efforts for designing a bigger network while avoiding overfitting, including drop out, weight regularization and batch normalization approaches. All these techniques gave reasonably good results in practice but still the main bottleneck for training bigger networks is to have access to enough number of training data which has been precisely labeled (for classification) or mapped (for data-fitting). In the proposed method the introduced architecture seems to avoid overfitting in comparison to the classical models with the same number of parameters.
2. The other big problem in making bigger networks is its computational cost. Training bigger networks need more resources and they become computationally prohibitive which is an inevitable fact. Training procedure of the proposed architecture shows faster convergence and also results in the less training loss in comparison to the classic models with the same number of parameters. This gives us the advantage of using smaller networks and less number of parameters while getting the same loss value which takes us one step closer to make an implementable network in costumer devices.

## SPDNN idea

Suppose there is a set of several neural networks designed for a specific task. The main idea of Semi Parallel Deep Neural Networks (SPDNN) is to compensate the drawbacks of each neural network using other networks in that set. For example, suppose that for a specific deep learning problem one can find N different successful networks in the literature which each of them gives "reasonable" results on that specific task. Each of these networks has their own drawbacks and would fail on some input data. The SPDNN idea uses a parallel version of these networks and merges all of them into a single network. Whenever results of two layers are going toward a single layer the concatenation technique is used to mix the path inside the network. And for the output layer (if sizes of the kernel are not same) the following technique is used to merge all layers. There could be two different possibilities in this step:

1. The output of each network is a p channel image: In this case, the N outputs of the N networks are placed in a N*p channel convolutional layer and a kernel by the shape (k1, k2, N*p, p) is placed after this layer to map it to a p channel image. Where (k1, k2) is the size of the kernel. See figure 1.
2. The output of each network is a vector with length p (classic classification or data fitting): In this case the outputs are concatenated to generate a neural layer of the size p*N and the weight matrix by the size (p*N, p) is mapped this layer to the output layer which has the length p. See figure 2.

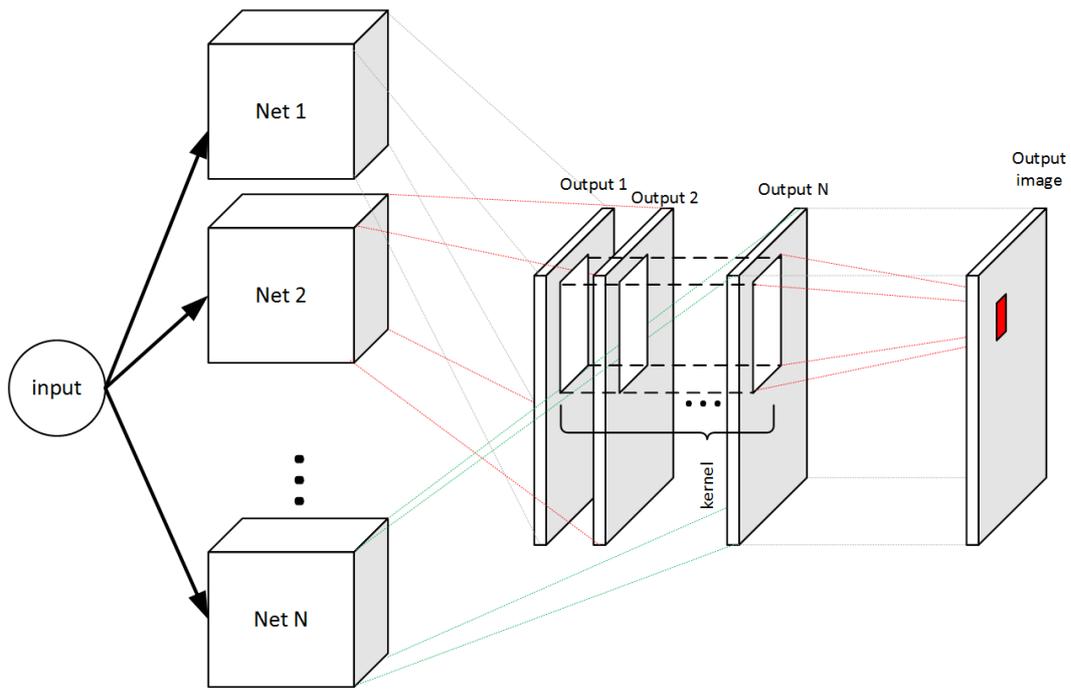

Figure 1: concatenating the last layer into a single layer (convolution case)

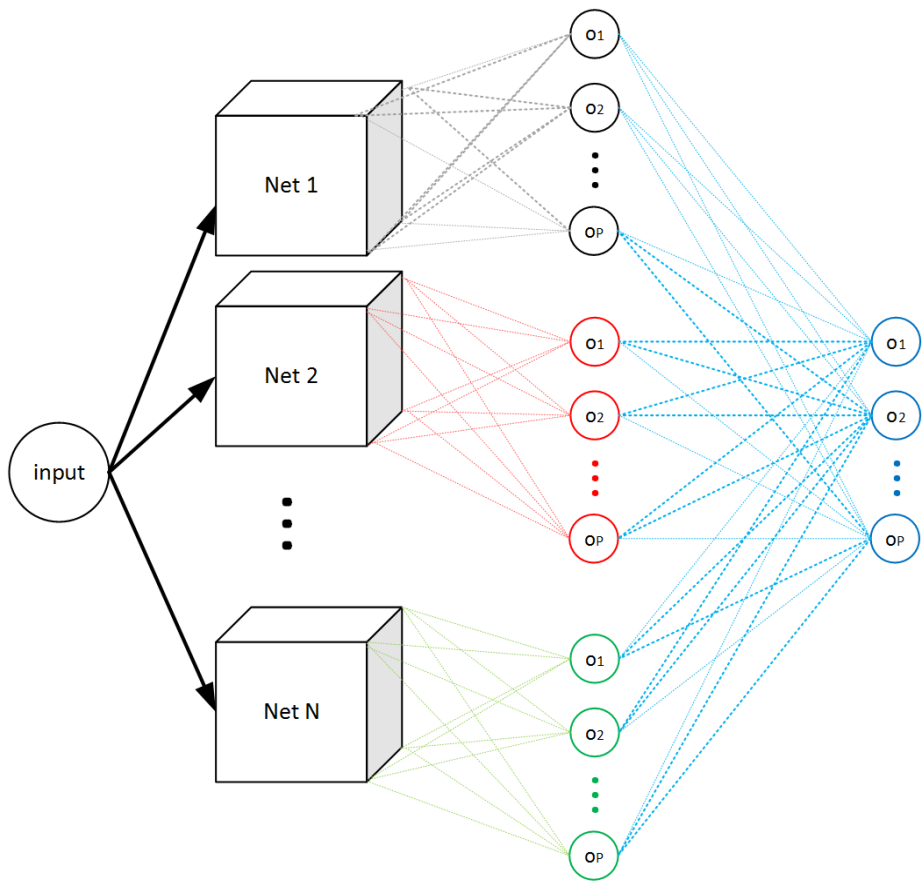

Figure 2: concatenating the last layer into a single layer (Fully connected case)

The SPDNN technique is described with an example in the following section. The importance of this architecture is in the backpropagation step of the training procedure:
1. In the feed forward step, the networks are not seeing each other. The data goes into each network and the output is calculated and their outputs are merged using one of the two different methods described above.
2. In the back propagation step, the loss function is calculated from the merged output of the networks, and while the error is back-propagated throughout the networks to the input, these N different networks placed in SPDNN scheme are getting influenced by each other.

## SPDNN with an example

The problem we are talking about here is the low-quality iris image segmentation. The extended CASIA 1000 dataset is used to train three different models. These three models have almost the same number of parameters.

1. The first network is an 8 layer fully convolutional deep neural network designed to the iris segmentation task. All layers are using 7x7 kernels. The Batch normalization technique is used after each convolutional layer to avoid overfitting and faster convergence.(Figure 3).

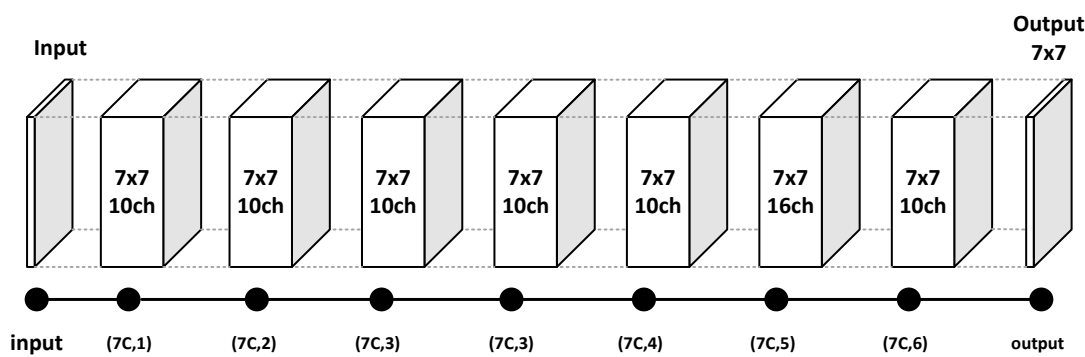

*Figure 3: network 1 used to perform iris segmentation.*

2. The second network designed for the problem in hand is a 6 layer fully convolutional network shown in Figure 4.The kernel size 3x3 has been used in all layers. No pooling is used in the network, and the batch normalization technique is used again after each convolutional layer.

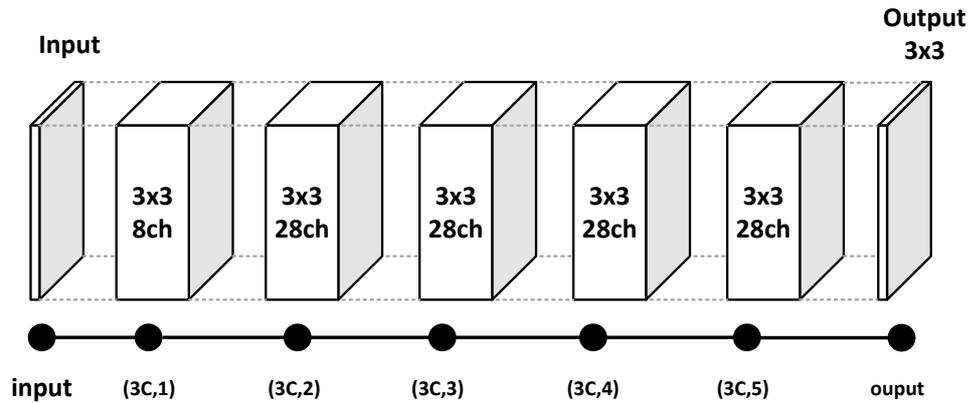

Figure 4: network 2 used to perform iris segmentation.

3. The third proposed network is a 5 layer fully convolutional neural network shown in Figure 5. The kernel size is increasing for each layer starting with 3x3 for the first layer and 11x11 for the output layer. No pooling is used in the network, and the batch normalization technique is used again after each convolutional layer.

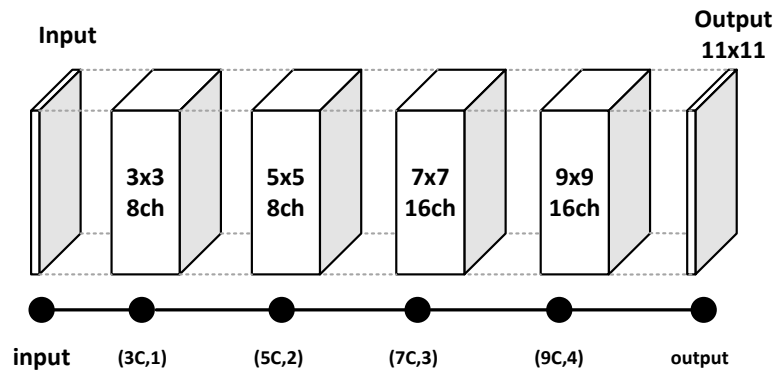

Figure 5: network 3 used to perform iris segmentation.

Three networks described above are trained using Nestrov Momentum method for Binary Cross-Entropy loss function on expanded CASIA 1000 dataset. The training loss and validation loss is shown in *Figure* 6 and *Figure* 7 respectively (the loss axis is in log scale).

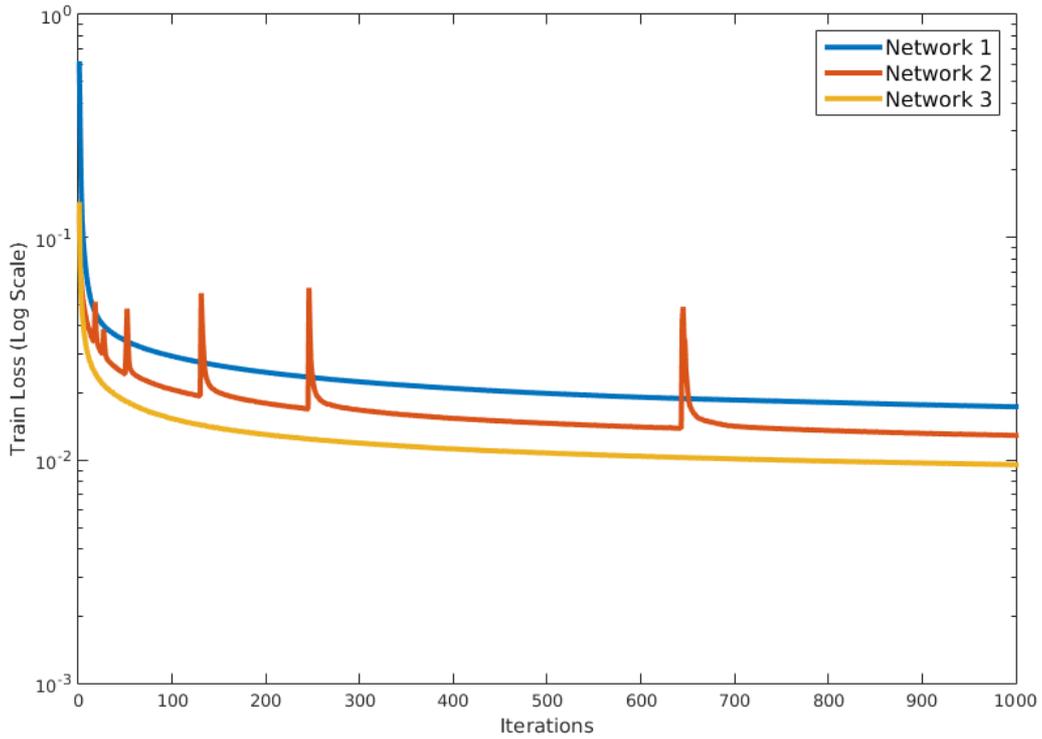

*Figure 6: Train loss for three networks*

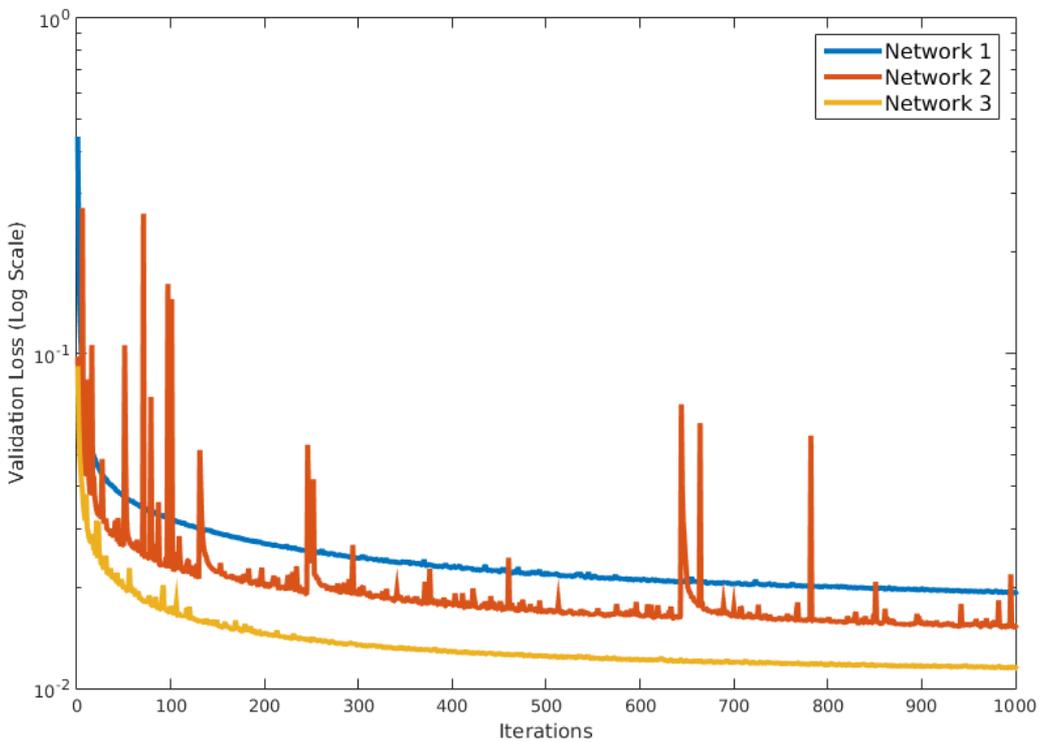

*Figure 7: Validation loss for three networks*

Testing is done for 20% of the samples in the CASIA 1000 which has not been seen by the network during the training procedure. The following statistical measure and performance have been calculated for the output of each network:

1. Accuracy (closer to 1 better)
2. Sensitivity (closer to 1 better)
3. Specificity (closer to 1 better)
4. Precision (closer to 1 better)
5. Negative Predictive Value (NPV) (closer to 1 better)
6. False Positive Rate (FPR) (closer to 0 better)
7. False Negative Rate (FNR) (closer to 0 better)
8. False Discovery Rate (FDR) (closer to 0 better)
9. F1 Score (closer to 1 better)
10. Matthew Correlation Coefficient (MCC) (closer to 1 better)
11. Informedness (closer to 1 better)
12. Markedness (closer to 1 better)

*Figure* 8 to *Figure* 10 are illustrating the histogram of the above measurements for each network in testing step. The values written on the top of each histogram is the mean value for that measurement.

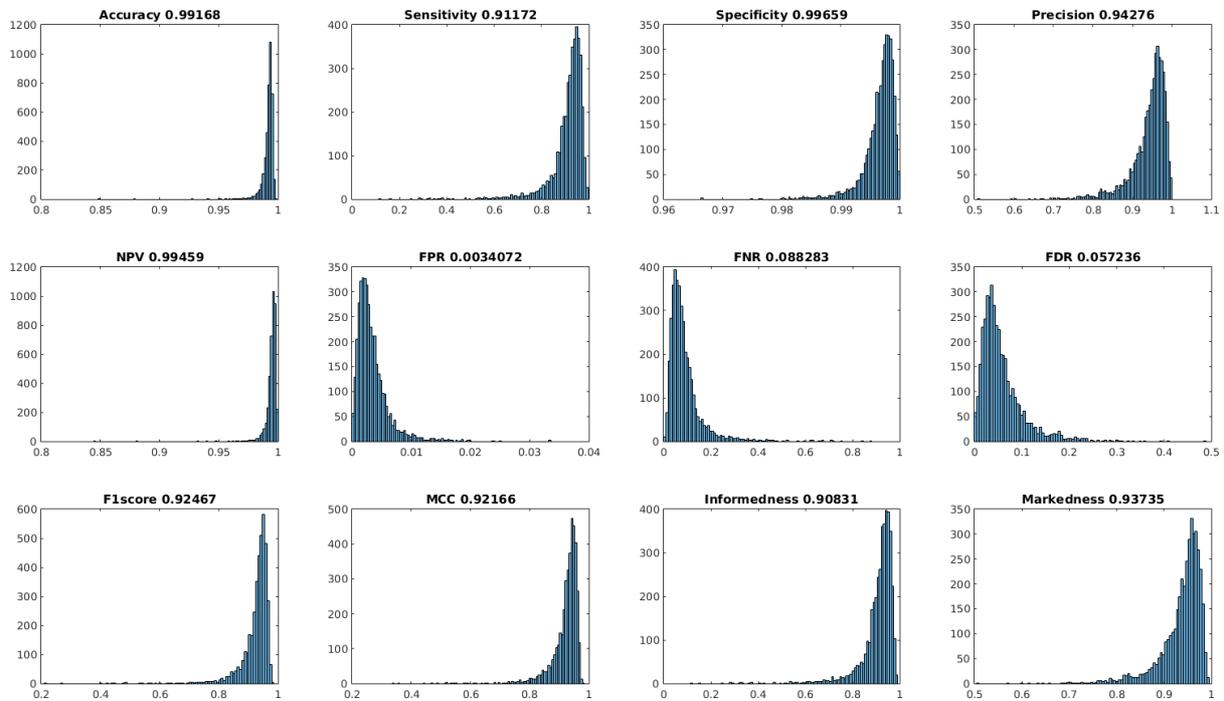

*Figure 8: Testing measurements on network 1*

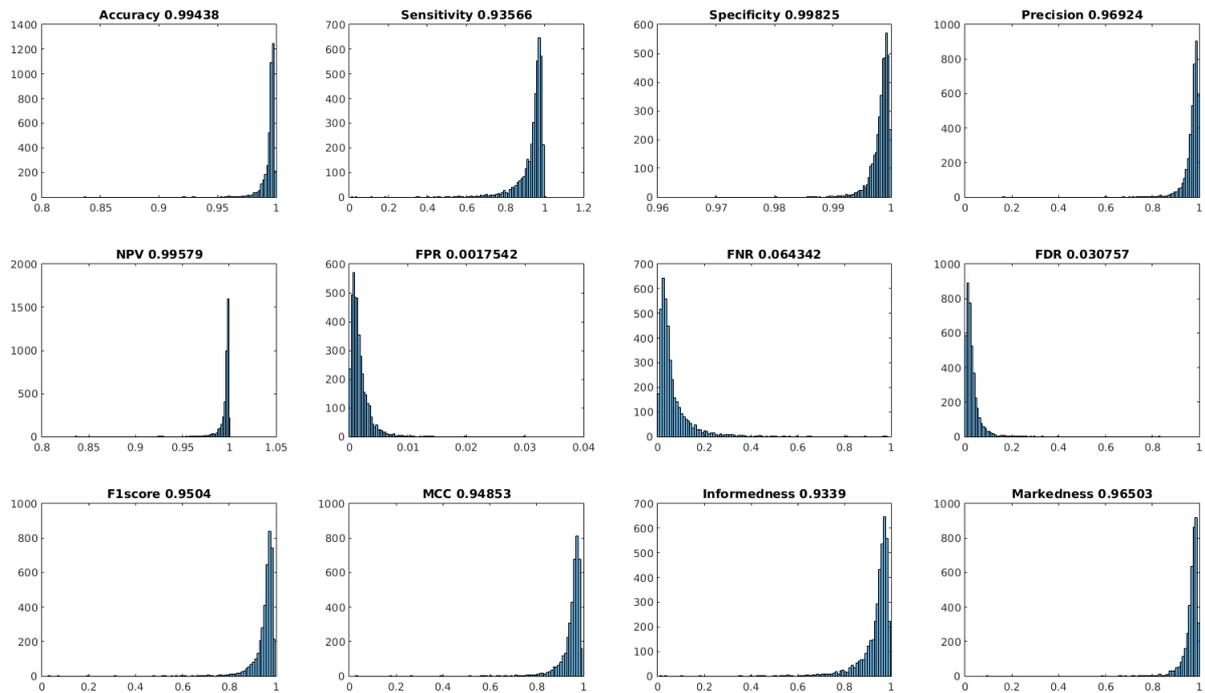
*Figure 9: Testing measurements on network 2*

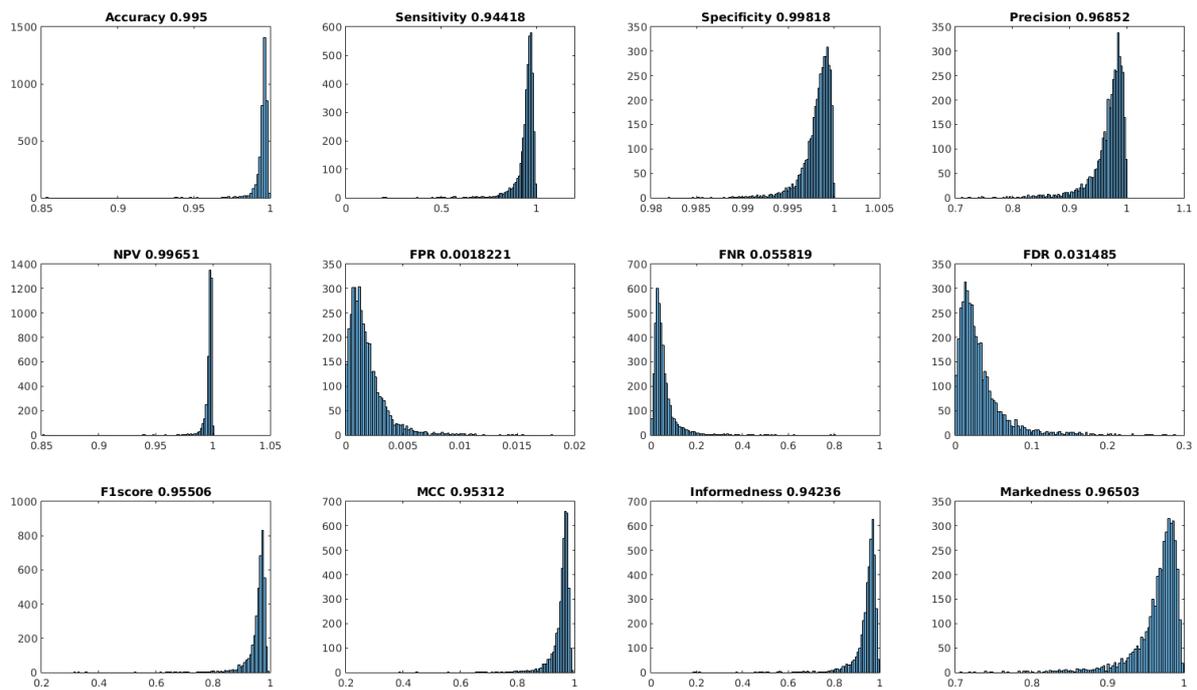
*Figure 10: Testing measurements on network 3*

## SPDNN Explained with an example:

The main purpose of the SPDNN model is to mix and merge several deep architectures and produce a final model which is taking advantage of all architectures. This is happening by translating each network to a graph, label each node of the graph based on the network architecture, apply the graph contraction to the network graph and translate back the graph into the final network. The method is described step by step below:

1- Translate each network into its corresponding graph. The graph for each of three networks proposed here are shown under the networks in figures 3 top 5. each layer of network corresponds to a node in the graph.
2- Assign properties to each node. Each node of the graph takes two properties. 1- first property is the operation of the layer. Letters used to label each operation, C for convolutional, F for fully connected layers and P for pooling operation. In this work, there are no fully connected and pooling layers. So just the C sign is used in all layers. For example, 5C correspond to a convolutional layer with the 5x5 kernel. 2- The second label is the distance of this layer to the input layer. For example, (9C,4) is the label of the node operating 9x9 convolutional operation and has distance 4 from the input node.
3- In this step all these graphs are placed in parallel, sharing same input and output. See Figure 11. All the nodes with the same properties are assigned to the same label. For examples, all the nodes with the property (7C,3) are labeled as C.

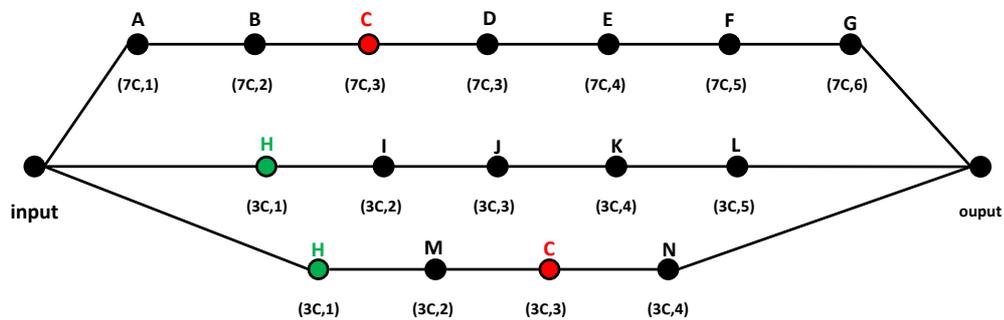

Figure 11: All graphs are placed in parallel sharing a single input and output

4- The graph contraction operation is applied to this graph. In graph contraction step all the nodes with the same label are merged into a single node while saving their connections to the previous and next nodes. See Figure 12

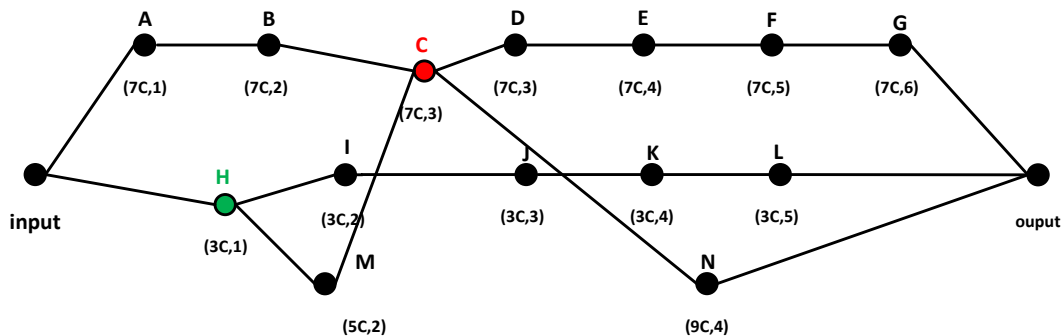

Figure 12: The graph contraction is applied to the graph in Figure 11

5- The last step is to translate the graph back into the neural network. In the nodes where two or more nodes are merging the concatenation operation is applied. And the operation for each node is specified by the first property of the node. The final network is shown in Figure 13

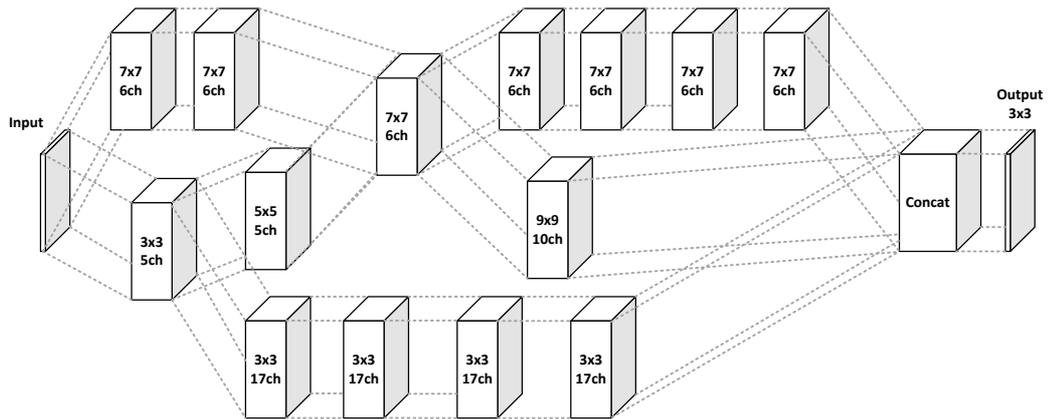

Figure 13: the contracted graph is translated into the neural network.

In order for the comparisons to be fair, the number of the channels in the final design is selected in a way for the network to have almost same number of parameters as each of its parent networks.

This model is trained using the same method (Nestrov Momentum) for the same loss function (Binary Cross-entropy). The training and validation losses during the training procedure are shown in *Figure 14* and *Figure 15* respectively.

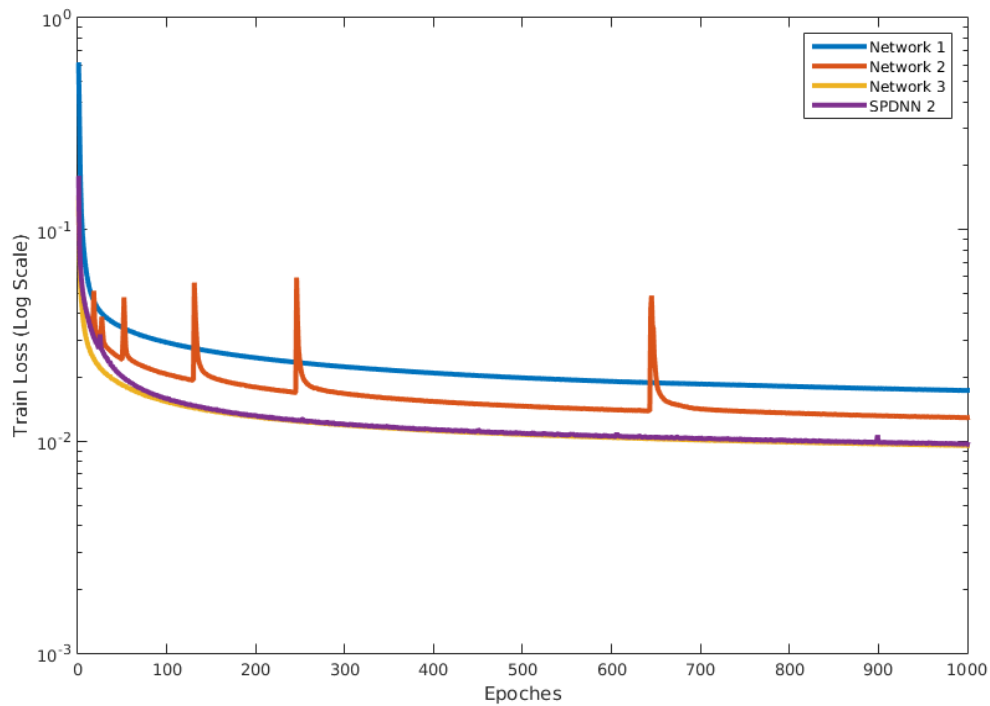

Figure 14: Train loss for all networks including SPDNN network. The network with SPDNN idea is labeled as SPDNN2.

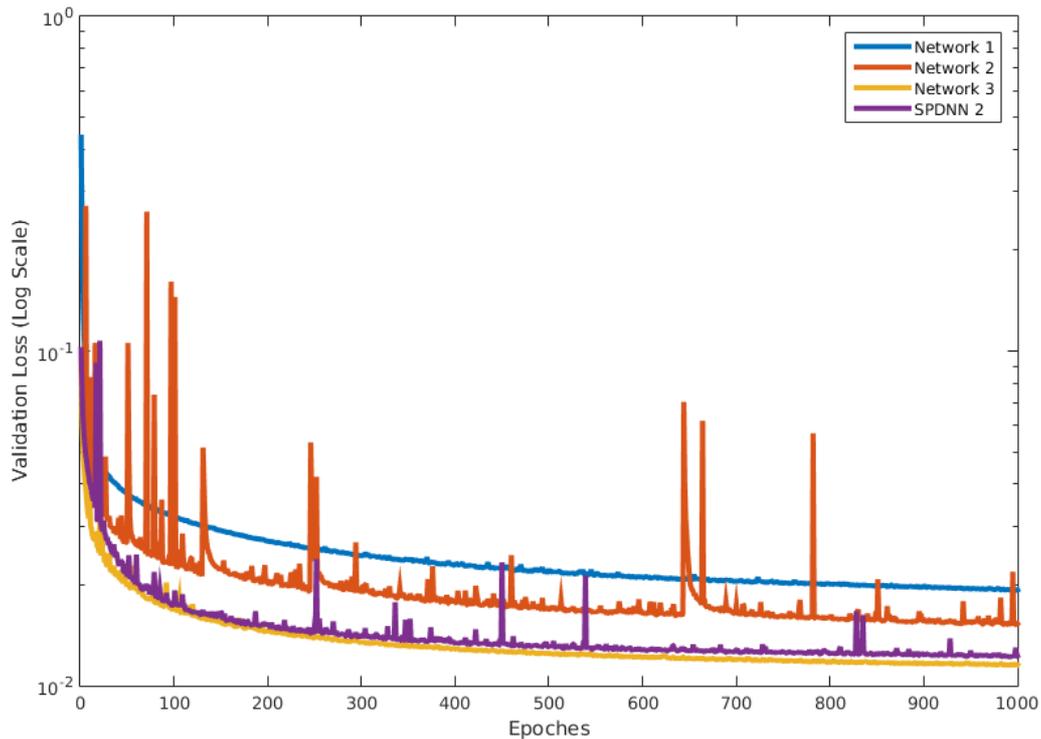

*Figure 15: Validation loss for all networks including SPDNN network. The network with SPDNN idea is labeled as SPDNN2.*

As you can see the losses converge to the best loss of the three networks. This shows that while having several networks using SPDNN helps the model to converge to the best available network. The testing results for the new semi-parallel network is given in *Figure 16*.

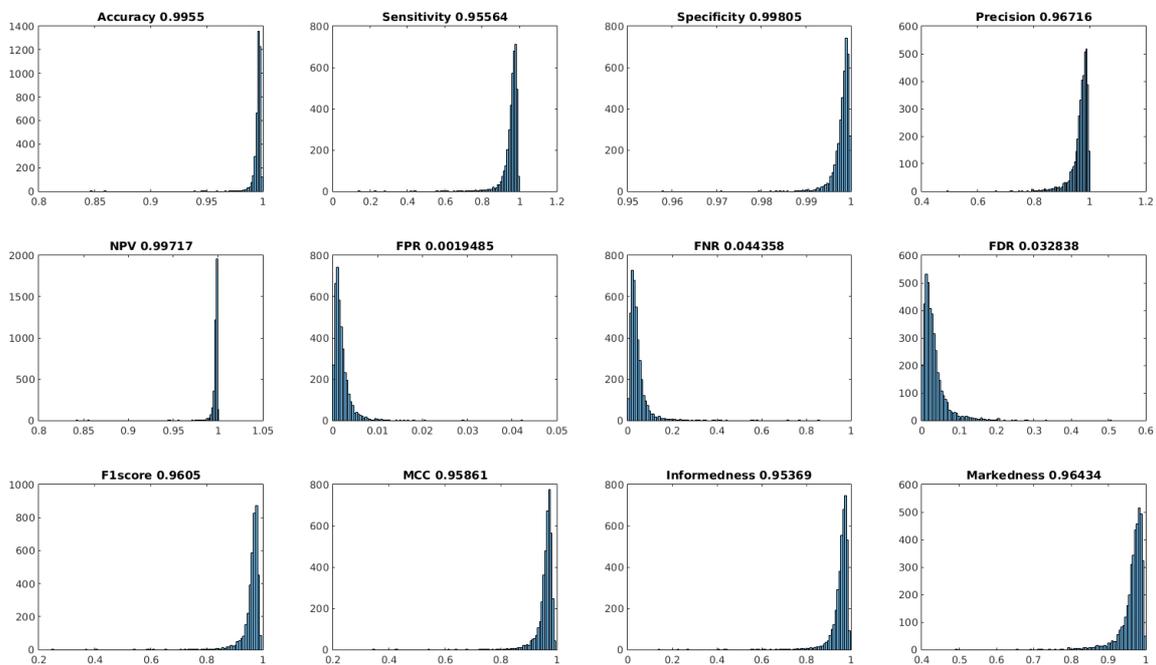

*Figure 16: Test results for SPDNN model.*

The comparisons on the test dataset for all networks are shown in *Figure 17* and *Figure 18*.

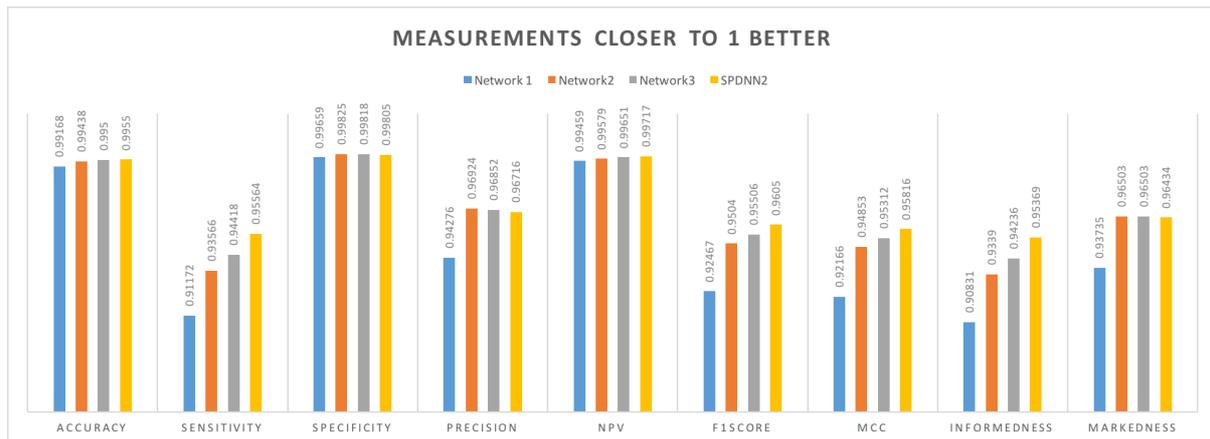
*Figure 17: Comparisons for all networks. Higher values indicate better performance.*

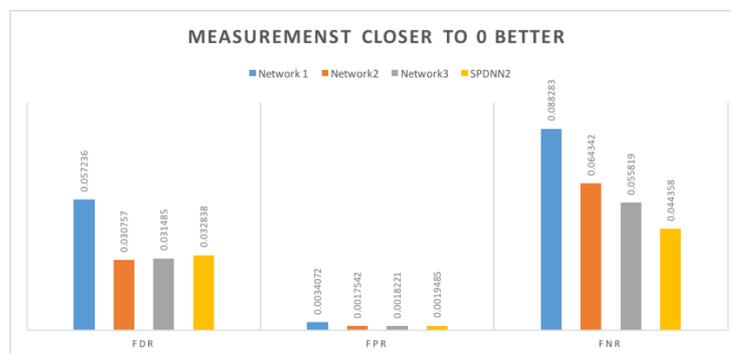
*Figure 18: Comparisons for all networks. Lower values indicate better performance.*

From figures 16 to 18 it is concluded that, although the SPDNN2 network has the same number of parameters as the networks 1, 2 or 3. It gives better results in most of the measurements including accuracy, sensitivity, NPV, F1Score, MCC, Informedness and FNR. The higher accuracy, F1-score, and MCC show that the SPDNN idea has a better performance, in general, compare to its parent networks. The measurement values that the SPDNN couldn't overtake the other networks are very close to the best network in the parallel set.

From the training and testing results we can see that the SPDNN helps the mixture of all networks to merge to the best design and it increases the convergence in the training stage and in the test part, it has been shown that the SPDNN can help to generalize the result more that each and every of parent networks.

## Similar ideas
**Ensemble Classifiers**

In the ensemble classifiers, a different set of classifiers are training for the problem individually and then they're placed in parallel in to form a bigger classifier and the results of the all of these classifiers are used to take the final decision [1]. The difference of SPDNN is that the networks placed in parallel are not trained individually. In fact, they are training together and as described before they are influenced by each other in the backpropagation step. From the other point of view, SPDNN can be used for data fitting as well, while the ensemble classifiers are not capable of.

**Google Inception Units**

As claimed by the authors in [2] the inception model is "**to consider how an optimal local sparse structure of a convolutional vision network can be approximated and covered by readily available dense components**". Actually Inception unit is the approximation of one and just one layer in the network. See *Figure 19*.

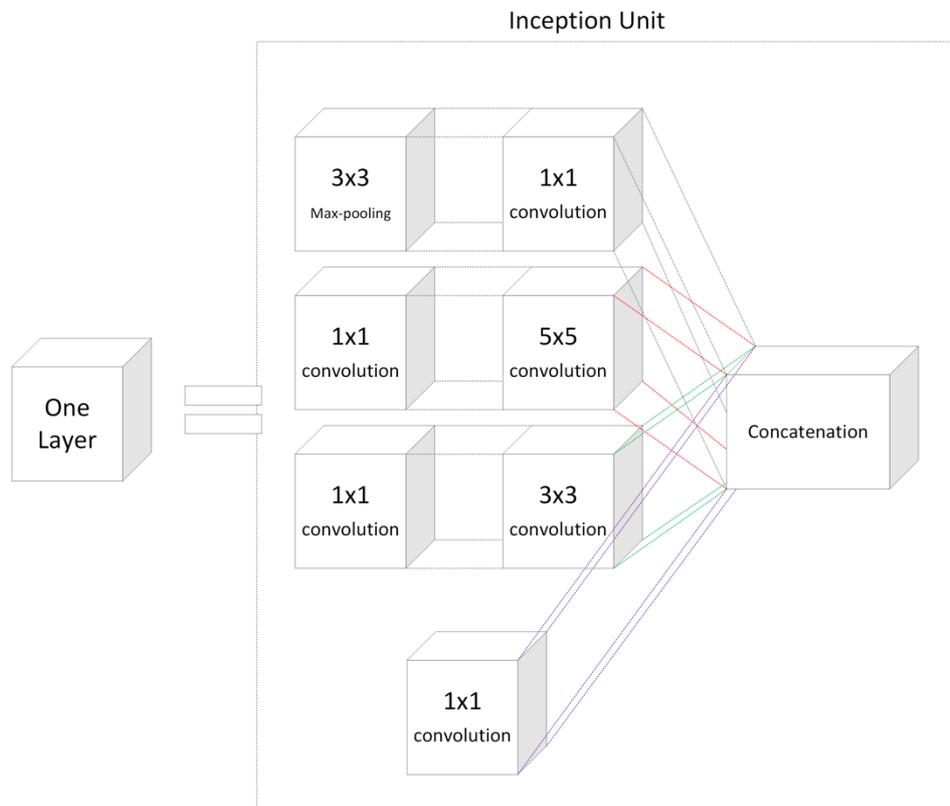

*Figure 19: inception unit*

1. In this idea, the Inception units are placed consecutively after each other to make the network deeper and deeper but in the SPDNN idea, different networks are placed in parallel in order to make the resulting network wider other than deeper. In fact, one of the different networks placed in the SPDNN design could be a network made by Inception units. This makes the SPDNN idea different from the Inception.
2. From the other side, as the authors claimed, the inception idea is designed for convolutional vision networks, where, as discussed before SPDNNs could be used for both convolutional and dense neural outputs.

## Acknowledgement

This research is funded under the SFI Strategic Partnership Program by Science Foundation Ireland (SFI) and FotoNation Ltd. Project ID: 13/SPP/I2868 on Next Generation Imaging for Smartphone and Embedded Platforms.
We gratefully acknowledge the support of NVIDIA Corporation with the donation of a Titan X GPU used for this research.
Portions of the research in this paper use the CASIA-IrisV4 collected by the Chinese Academy of Sciences' Institute of Automation (CASIA).